\documentclass[sigconf]{acmart}

\usepackage{multirow}

\AtBeginDocument{%
  }

\copyrightyear{2025}
\acmYear{2025}
\setcopyright{acmlicensed}\acmConference[ICMR '25]{Proceedings of the 2025 International Conference on Multimedia Retrieval}{June 30-July 3, 2025}{Chicago, IL, USA}
\acmBooktitle{Proceedings of the 2025 International Conference on Multimedia Retrieval (ICMR '25), June 30-July 3, 2025, Chicago, IL, USA}
\acmDOI{10.1145/3731715.3733471}
\acmISBN{979-8-4007-1877-9/2025/06}

\begin{document}

\title{Bridging the Gap Between Semantic and User Preference Spaces for Multi-modal Music Representation Learning}

\author{
    Xiaofeng Pan \footnotemark[1] \\
    Jing Chen
}
\authornote{The first two authors contributed equally to this paper.}
\email{
    pxfvintage@163.com
}
\email{
    thunsir@gmail.com
}
\affiliation{
  \institution{NetEase Inc.}
  \country{Hangzhou, China}
}

\author{
    Haitong Zhang \\
    Menglin Xing \\
    Jiayi Wei
}
\email{
    zhanghaitong03@corp.netease.com
}
\affiliation{
  \institution{NetEase Inc.}
  \country{Hangzhou, China}
}

\author{
    Xuefeng Mu \\
    Zhongqian Xie
}
\email{
    hzmuxuefeng@corp.netease.com
}
\affiliation{
  \institution{NetEase Inc.}
  \country{Hangzhou, China}
}

\renewcommand{\shortauthors}{Xiaofeng Pan et al.}

\begin{abstract}
Recent works of music representation learning mainly focus on learning acoustic music representations with unlabeled audios or further attempt to acquire multi-modal music representations with scarce annotated audio-text pairs. They either ignore the language semantics or rely on labeled audio datasets that are difficult and expensive to create. Moreover, merely modeling semantic space usually fails to achieve satisfactory performance on music recommendation tasks since the user preference space is ignored. In this paper, we propose a novel \textbf{H}ierarchical \textbf{T}wo-stage \textbf{C}ontrastive \textbf{L}earning (HTCL) method that models similarity from the semantic perspective to the user perspective hierarchically to learn a comprehensive music representation bridging the gap between semantic and user preference spaces. We devise a scalable audio encoder and leverage a pre-trained BERT model as the text encoder to learn audio-text semantics via large-scale contrastive pre-training. 
Further, we explore a simple yet effective way to exploit interaction data from our online music platform to adapt the semantic space to user preference space via contrastive fine-tuning, which differs from previous works that follow the idea of collaborative filtering. As a result, we obtain a powerful audio encoder that not only distills language semantics from the text encoder but also models similarity in user preference space with the integrity of semantic space preserved. 
Experimental results on both music semantic and recommendation tasks confirm the effectiveness of our method.
\end{abstract}




\maketitle

\section{Introduction}
Music representations that combine multi-modal song information would help achieve high performance across various downstream tasks \cite{shen2023more, stoikos2023cross, yang2024beatdance}. 
With the rapid progress of deep learning, self-supervised learning methods (e.g., MERT \cite{yizhi2023mert} and Audio-MAE \cite{huang2022masked}) and contrastive learning methods (e.g., COLA \cite{saeed2021contrastive} and CLMR \cite{spijkervet2021contrastive}) have been proven effective in learning meaningful representations with unlabeled audios. Further, CLAP \cite{elizalde2023clap} proposes to learn audio concepts from natural language supervision. 
It requires labeled audio datasets which are difficult, expensive and time-consuming to create for music in particular due to the high technicality and subjectivity. All these methods overlook the rich semantic information conveyed in textual music metadata, especially lyrics, which deliver both linguistic and musical messages in the form of natural language \cite{watanabe2020lyrics} and play a key role in music understanding.

Besides, interaction data between users and songs, a new modality of music, are available in vast quantities but usually ignored in general-purpose music representation learning. As demonstrated in \cite{yang2023courier}, simply incorporating semantic representations into search and recommendation tasks yields only marginal improvements. Therefore, several methods \cite{ferraro2021enriched, zhao2023bootstrapping} propose to learn music representations by combining the above all modalities with contrastive learning. They follow the idea of collaborative filtering, assuming songs that co-occur in a user behavior sequence or a human-created playlist to be similar. However, this assumption is flawed since user behaviors usually carry heavy noise and have no inherent connection to the key notion of similarity in contrastive learning. Guiding the model towards an uncertain similarity metric is likely to bring conflicts with semantic similarity modeling, e.g., songs not similar in semantic space may easily co-occur in user behaviors and therefore be considered as "similar".

Based on these observations, we propose a novel \textbf{H}ierarchical \textbf{T}wo-stage \textbf{C}ontrastive \textbf{L}earning (HTCL) method that models similarity from the semantic perspective (i.e., audio and text) to the user perspective (i.e., song and song) hierarchically to bridge the gap between semantic and user preference spaces. 
In contrast to previous works (e.g., MERT, Audio-MAE, CLAP) that train on audio clips and aggregate clip-level representations for full audio representations, we directly model full-length music audio for simplicity, efficiency, and representation integrity.

To be specific, we leverage a pre-trained BERT model to encode text information of a song and integrate open-world knowledge. Meanwhile, a scalable audio encoder is devised based on the convolutional neural network (CNN) \cite{li2021survey} and Transformer \cite{vaswani2017attention} structure, allowing for flexible handling of computational and storage issues. In the first stage, we sample songs from our music platform (60 million DAU) and organize them into audio-text pairs to build a large-scale contrastive pre-training dataset, enabling the learning of multi-modal semantics and relationships between them. In the second stage, we utilize interaction logs from the Similar Recommendation Channel in our platform where users can see the trigger song of each recommended song, so we can collect music pairs that users assume to be similar and construct a dataset with millions of <trig\_audio, rec\_audio, rec\_text> triplets for further contrastive fine-tuning. We believe that similarity voted by users is much more solid than hypothetical "similarity" mined by strategies in existing works \cite{saeed2021contrastive, spijkervet2021contrastive, ferraro2021enriched, zhao2023bootstrapping}. In this way, fewer conflicts are introduced into semantic space when modeling user-preferred similarity.

Our contributions can be summarized as follows:

\noindent $\bullet$ To the extent of our knowledge, this is the first study of multi-modal music representation learning paying attention to the gap between semantic and user preference spaces. We propose a simple yet effective method named HTCL, which models similarity from semantic to user perspectives hierarchically to bridge the gap.

\noindent $\bullet$ We devise a scalable audio encoder to distill language semantics from a pre-trained BERT model, learning audio-text semantics via large-scale contrastive pre-training. Further, we propose an effective approach to exploit interaction data via contrastive fine-tuning to adapt the semantic space to user preference space.

\noindent $\bullet$ Our method is evaluated on real-world datasets from our platform and outperforms state-of-the-art (SOTA) approaches on both music semantic and recommendation tasks. Further, we conduct extensive analysis to confirm the effectiveness of our design. The code and part of evaluation datasets is publicly available\footnote{https://github.com/AaronPanXiaoFeng/HTCL} to facilitate reproducibility and further research.

\section{Proposed Method} \label{sec:method}

\begin{figure*}[htbp]
    \setlength{\abovecaptionskip}{-0.1cm}
    \setlength{\belowcaptionskip}{-0.4cm}
    \centering
    \includegraphics[width=1\linewidth]{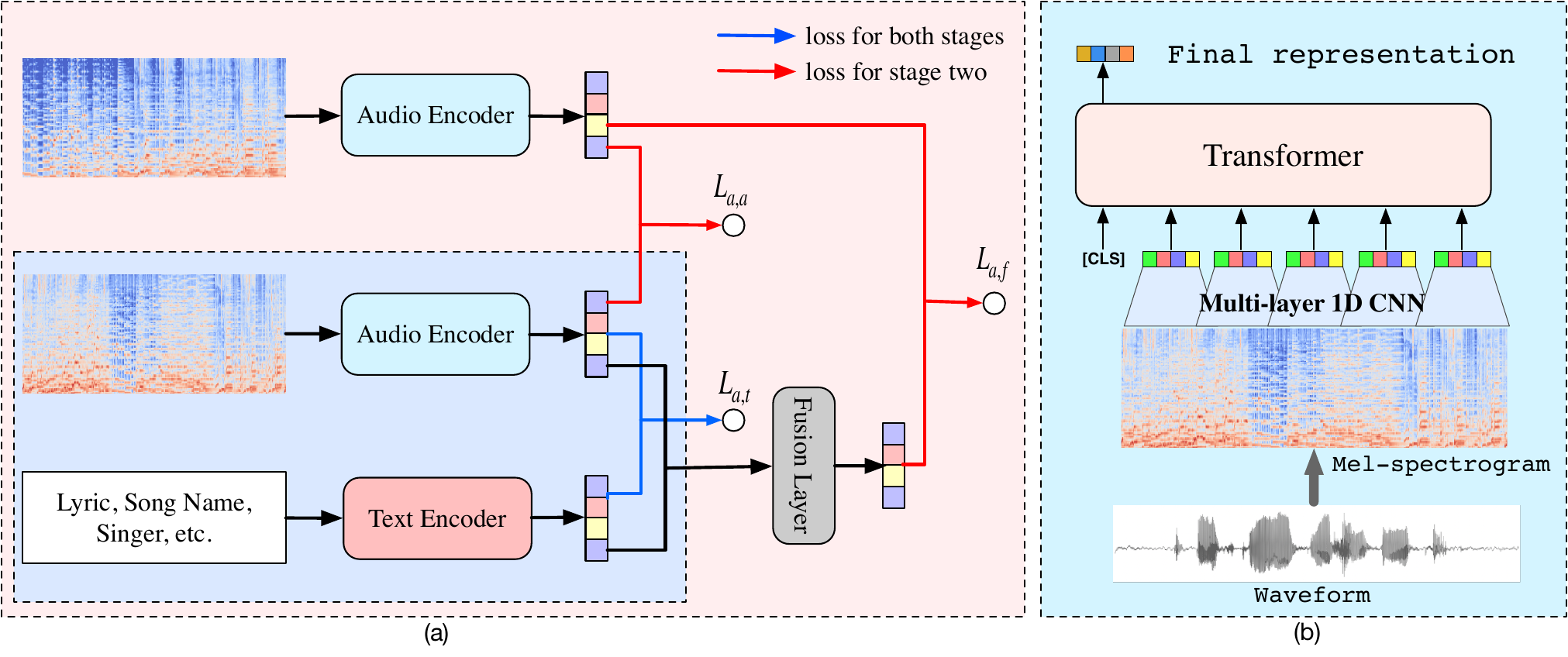}
    \caption{(a) Overall framework of HTCL: two-stage contrastive learning. (b) Model structure of the audio encoder.}
    \label{fig:arch}
\end{figure*}

As shown in Figure~\ref{fig:arch}, the first stage of our HTCL focuses on modeling semantic similarity from a lower-level perspective (i.e., audio and text), while the second stage attends to a higher-level perspective (i.e., song-to-song relationships) to capture user-preferred similarity. The detailed model design will be presented in the following sections.

\subsection{Contrastive Pre-training}
For contrastive learning, we extract the first two minutes of audio (padding shorter tracks with silence) as $\mathbf{X}_a$ and concatenate textual music metadata (e.g., title, lyrics, artists) in a simple format as $\mathbf{X}_t$, creating abundant training pairs.

The HTCL mainly consists of two encoders, i.e., a text encoder $\mathcal{E}_{text}$ and an audio encoder $\mathcal{E}_{audio}$. We utilize a pre-trained BERT\footnote{https://huggingface.co/google-bert/bert-base-multilingual-uncased} model as text encoder to combine open-world knowledge and alleviate training difficulties. It aims to capture the rich linguistic and musical messages embedded in $\mathbf{X}_t$. Meanwhile, we design $\mathcal{E}_{audio}$ to extract meaningful features from $\mathbf{X}_a$. Due to the large size of audio files, directly using raw waveforms as inputs would result in substantial storage and computational costs, making it impractical to handle tens of millions of training samples. Therefore, we transform $\mathbf{X}_a$ into Mel-spectrograms \cite{gong2021ast} $\mathbf{x}_{mel}$ with a $T_h$ ms Hanning window that shifts every $T_s$ ms, which aligns better with the perception of human auditory system toward audios. Then we adopt the Transformer as backbone to capture the local and global patterns in the audio signals. Since the computational complexity of Transformer is proportional to the square of the input sequence length, we use a multi-layer 1-dimensional CNN to further compress $\mathbf{x}_{mel}$ with minimal information loss before it is fed into the Transformer. In this way, both efficiency and scalability can be achieved. Finally, we obtain semantic representations for downstream tasks as follows:
\begin{equation}
    \begin{aligned}
        \mathbf{z}_t &= \mathcal{E}_{text}(\mathbf{X}_{t}) \in \mathbb{R}^D \\
        \mathbf{z}_a &= \mathcal{E}_{audio}(\mathbf{x}_{mel}) \in \mathbb{R}^D
    \end{aligned}
\end{equation}
where $\mathbf{z}_t$ and $\mathbf{z}_a$ denotes the textual representation and acoustic representation, respectively.

For a batch with $B$ training samples, we formulate the objective of contrastive pre-training as follows:
\begin{align}
    & L_{a \rightarrow t} = -\frac{1}{B}\sum_{i=1}^B \log\frac{\exp(\mathbf{z}_a^{(i)} \cdot \mathbf{z}_t^{(i)}/\tau)}{\sum_{j=1}^B \exp(\mathbf{z}_a^{(i)}\cdot\mathbf{z}_t^{(j)}/\tau)} \\
    & L_{t \rightarrow a} = -\frac{1}{B}\sum_{i=1}^B \log\frac{\exp(\mathbf{z}_t^{(i)} \cdot \mathbf{z}_a^{(i)}/\tau)}{\sum_{j=1}^B \exp(\mathbf{z}_t^{(i)}\cdot\mathbf{z}_a^{(j)}/\tau)} \\
    & L_{a,t} = L_{a \rightarrow t} + L_{t \rightarrow a}
\end{align}
where $\cdot$ refers to inner product and $\tau$ is a temperature parameter to scale the range of logits. $L_{a \rightarrow t}$ and $L_{t \rightarrow a}$ measure the similarity between audios and texts from opposite directions. We use the symmetric loss function $L_{a,t}$ to jointly train the audio encoder and the text encoder, with a small learning rate applied to $\mathcal{E}_{text}$ to tackle the catastrophic forgetting problem.

\subsection{Contrastive Fine-tuning}
In the second stage, we aim to adapt the learned semantics to the user preference space with the integrity of semantic space preserved. To achieve this, we explore a simple yet effective way to exploit interaction data between users and songs. 

Different from existing works \cite{ferraro2021enriched, zhao2023bootstrapping} that generate co-occurring song pairs for contrastive learning, we build training samples based on user favored songs in the Similar Recommendation Channel where users can see the trigger song of each recommended song. We consider that the similarity voted (i.e., favored) by users is more solid, since co-occurring song pairs mined from user behaviors usually carry heavy noise and are not inherently similar.

Specifically, we construct a dataset with millions of <trig\_audio, rec\_audio, rec\_text> triplets and attempt to minimize the distance between representations of the recommended song and its trigger audio. With the pre-trained $\mathcal{E}_{text}$ and $\mathcal{E}_{audio}$, we obtain representations for the trigger audio, recommended audio and textual music metadata, i.e., $\mathbf{z}_a^{T}$, $\mathbf{z}_a^{R}$ and $\mathbf{z}_t^{R}$. The $\mathbf{z}_a^{R}$ and $\mathbf{z}_t^{R}$ are concatenated and fed into a MLP (i.e., the Fusion Layer) to compute $\mathbf{z}_f^{R}$, i.e., representation of the fused semantics. For contrastive fine-tuning, we formulate training losses as follows:
\begin{equation} \label{eq:sft_loss}
    \begin{aligned}
        L_{a,a} &= L_{a^T \rightarrow a^R} + L_{a^R \rightarrow a^T} \\
        L_{a,f} &= L_{a^T \rightarrow f^R} + L_{f^R \rightarrow a^T} 
    \end{aligned}
\end{equation}
where $a^T$, $a^R$ and $f^R$ refer to the trigger audio, the recommended audio and the fused semantics, respectively. $L_{a,a}$ and $L_{a,f}$ are devised to align the trigger audio with the recommended audio and the trigger audio with the fused semantics, respectively. Each term in the right of Equation~\ref{eq:sft_loss} can be calculated in a similar way with Equation 2 and 3. Besides, $L_{a,t}$ is calculated to strengthen the relationships between audio and text modalities of the recommended song, which is considered as a strong positive sample for contrastive learning as it's favored by users. By minimizing the sum of $L_{a,a}$, $L_{a,f}$ and $L_{a,t}$, we develop a powerful audio encoder that not only distills language semantics from the text encoder but also models user-preferred similarity.

\section{Experiments}

\subsection {Experimental Setup}

\subsubsection{Datasets}
\label{subsec:datasets}
We establish five datasets for the two-stage training and the downstream evaluations.

\noindent $\bullet$ \textbf{Pre-training and fine-tuning.} For the first-stage contrastive pre-training, we build a 50-million-song dataset from our platform's music library with quality and diversity considered. For the second-stage contrastive fine-tuning, we collect favor behaviors from the Similar Recommendation Channel where users can see the trigger song of each recommended song, building a dataset with approximately 4 million triplets of <trig\_audio, rec\_audio, rec\_text> through deduplication and sampling. The detailed statistics are shown in Table~\ref{tab:datasets}.

\noindent $\bullet$ \textbf{Music Semantic Tasks.} We evaluate music representations in semantic space using two fundamental tasks from our business scenario: music genre and language classification. Existing public datasets mainly consist of audio clips, whereas our goal is to classify the full music audio. Therefore, we construct a dataset consisting of 30k full songs, sampled evenly across 24 genres and 26 languages to ensure a comprehensive evaluation. The dataset will be made publicly available for future research.

\noindent $\bullet$ \textbf{Recommendation Tasks.} We evaluate music representations in user preference space during both matching and ranking stages of music recommendation. The matching dataset is derived from 3-month logs of our primary recommendation scenario, while the ranking dataset is built from 8-day logs. The detailed statistics are shown in Table~\ref{tab:datasets}.

\subsubsection{Competitors}
\textbf{MERT} \cite{yizhi2023mert} and \textbf{Audio-MAE} \cite{huang2022masked} are representative SOTA methods focusing on acoustic representation learning, while \textbf{CLAP} \cite{elizalde2023clap} learns multi-modal representations from audio-text pairs. Additionally, we developed two variants of HTCL, i.e., \textbf{HTCL\_w\_CF} and \textbf{HTCL\_w/o\_text}. Rather than using the dataset described in Section~\ref{subsec:datasets}, HTCL\_w\_CF follows the idea of collaborative filtering to generate co-occurring song pairs \cite{ferraro2021enriched, zhao2023bootstrapping} for contrastive fine-tuning. HTCL\_w/o\_text differs from HTCL by excluding the text modality during contrastive fine-tuning. To ensure fair comparison, we conduct experiments using representations obtained from the audio encoders of various methods.

\subsubsection{Implementation Details and Evaluation Metrics}
We transform raw waveform into 128-dimensional Mel-spectrograms with 128ms Hanning window that shifts every 96ms, resulting in 1x1251x128 feature map for a 120-second song. The audio encoder of our HTCL contains a 3-layer 1-dimensional CNN that has 512 channels with strides (2,2,2) and kernel widths (5,3,3), following by a 12-layer Transformer using 12-head attention with hidden\_size=768. 
We perform contrastive pre-training on 8 A100-80G GPUs (batch size: 800, learning rate: 1e-3) and fine-tuning on 3 A100-80G GPUs (batch size: 240, learning rate: 1e-4). To preserve open-world knowledge and mitigate catastrophic forgetting, we apply a small learning rate (3e-5) to the pre-trained BERT. The Adam optimizer is utilized throughout.
To ensure fair comparison, Audio-MAE and CLAP are also trained on our pre-training dataset following their empirically optimal hyperparameter settings.

For music semantic tasks, we feed music representations into a single-layer MLP for genre and language classification, using accuracy (\textbf{ACC}) as evaluation metric.
For matching stage of music recommendation, we choose next item prediction as evaluation task. For each user, we randomly select a favored song as the target and collect 30 previously favored songs as triggers. We then retrieve the top 10 similar songs for each trigger based on the cosine distance between music representations, using hit rate (\textbf{HR@100}) as evaluation metric.
For ranking stage of music recommendation, we predict click-through and favor probabilities (i.e., music CTR and CVR tasks), using \textbf{AUC} as evaluation metric. We train DIN \cite{DIN} models on the first 7 days of ranking dataset and evaluate on the last day, with embeddings initialized by music representations.

\begin{table}[t]
\centering
    \caption{Statistics of the established datasets}
    \vspace{-1em}
    \setlength{\tabcolsep}{1.1mm}
    {
        \begin{tabular}{l c c c c c}
            \hline
            Dataset &\#Users &\#Songs &\#Favors &\#Clicks &\#Impressions \\
            \hline
            Fine-tune & 4643944 & 573014 & 4097802 & - & - \\
            Matching & 336506 &	316115 & 18171629 & - & - \\
            Ranking & 2391820 & 434399 & 1867872 & 47540017 & 92786435 \\
            \hline
        \end{tabular}
    }
    \label{tab:datasets}
    \vspace{-1em}
\end{table}

\begin{table}[tp] 
\centering
    \caption{Results of comparison experiments}
    \vspace{-1em}
    \setlength{\tabcolsep}{1mm}
    {
    \begin{tabular}{l c c c c c c}
    \hline
    \multirow{2}{*}{Model} & \multicolumn{2}{c}{ACC} & HR@100 & \multicolumn{2}{c}{AUC} \\
    \cmidrule(r){2-3} \cmidrule(r){4-4} \cmidrule(r){5-6}
    & Genre & Language & Matching & CTR & CVR \\
    \hline
    MERT & 0.3937 & 0.3132 & - & - & - \\
    Audio-MAE & 0.4193 & 0.4117 & 1.92\% & 0.7103 & 0.7728 \\
    CLAP & \underline{0.4592} & \underline{0.6592} & 3.23\% & 0.7130 & 0.7745 \\
    HTCL\_w\_CF & 0.3739 & 0.4832 & \underline{10.57}\% & \underline{0.7133} & \underline{0.7760} \\
    HTCL\_w/o\_text & 0.4316 & 0.5632 & 10.14\% & - & - \\
    \textbf{HTCL} & \textbf{0.4616} & \textbf{0.6687} & \textbf{12.02\%} & \textbf{0.7257} &\textbf{0.7875} \\
    \hline
    \end{tabular}
    }
    \label{tab:experiments}
    \vspace{-1.5em}
\end{table}

\subsection{Experimental Results}
\label{subsec:exps}
The comparison results are presented in Table~\ref{tab:experiments} and the major observations are summarized as follows.

\noindent $\bullet$ Across both music semantic and recommendation tasks, CLAP outperforms SOTA acoustic representation methods MERT and Audio-MAE, particularly in language classification, emphasizing the importance of the text modality. 
However, the gap between semantic and user preference spaces results in limited performance gains when applying CLAP's multi-modal representations to music recommendation, particularly in ranking tasks.

\noindent $\bullet$ HTCL\_w\_CF attempts to bridge this gap through an additional contrastive fine-tuning stage. However, it performs poorly on music semantic tasks, 
as its fine-tuning stage assumes co-occurring songs in user behaviors to be similar, potentially conflicting with semantic similarity modeling.
Moreover, it fails to achieve satisfactory improvements in ranking tasks, as what it learns is highly homogenized with ranking models trained on user behaviors.

\noindent $\bullet$ HTCL yields the best performance across all tasks, achieving further improvements in music semantic tasks and significantly outperforming the runner-up in music recommendation. These results confirm the impacts of the gap between semantic and user preference spaces, and validate the effectiveness of our proposed method. Besides, HTCL\_w/o\_text underperforms CLAP in music semantic tasks, further emphasizing the importance of the text modality, even during contrastive fine-tuning. 

\subsection {Effectiveness analysis}

\begin{figure}[t]
    \vspace{-0.5em}
    \setlength{\abovecaptionskip}{-0.1cm}
    \centering
    \includegraphics[width=1\linewidth]{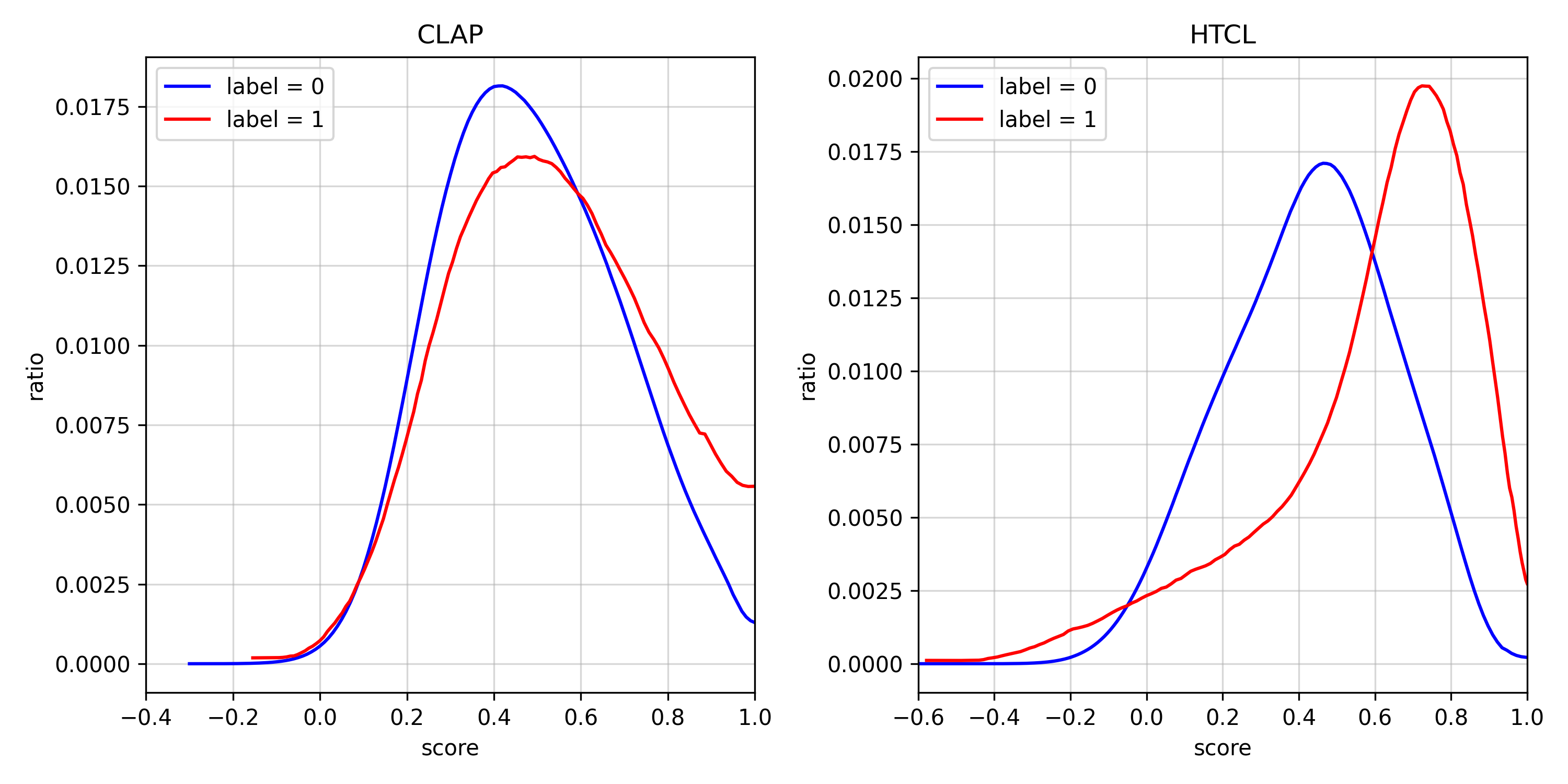}
    \caption{Distance distributions of anchor-positive and anchor-negative pairs calculated with representations from different methods.}
    \label{fig:distri}
    \vspace{-1.5em}
\end{figure}

Taking a step further, we investigate whether our approach can learn user-preferred similarity on the basis of semantic similarity, which we consider as the main reason for performance gains.
First, we randomly sampled 100k positive and negative samples respectively from 3-day logs of our Heuristic Search Channel where each sample is a song recommended based on a user's currently searched song (defined as the anchor). Therefore, positive samples, i.e., favored recommendations with corresponding anchors, can serve as good indicators of user-preferred similarity. We then computed the cosine distance score between representations of each sample and its anchor. 
As shown in Figure~\ref{fig:distri}, for HTCL, the distance distribution of anchor-positive pairs noticeably shifts to the right compared to that of anchor-negative pairs, indicating that user-favored recommended songs tend to have higher similarity scores with their corresponding anchors. In contrast, for CLAP, distance distributions of anchor-positive and anchor-negative pairs are barely distinguishable.

This visualization empirically demonstrates that HTCL's representations capture user-preferred similarity more effectively than prior methods.
Besides, with large-scale and diverse song collections utilized during contrastive pre-training, we can effectively guide the learning of HTCL through fine-tuning on a relatively small yet high-quality set of similar song pairs, while maintaining robust generalization capabilities.

\section{Conclusion}
In this paper, we investigate the challenges for multi-modal music representations and pay attention to the gap between semantic and user preference spaces. We propose a novel method named HTCL to combine modalities of audio, text, and user interactions to bridge the gap. By modeling similarity hierarchically in semantic and user preference spaces, our HTCL achieves better flexibility and generalization to downstream tasks. Future work will enhance the Fusion Layer via advanced multi-modal fusion techniques (replacing the simple MLP) and optimize music recommendation with multi-modal representations.

\end{document}